\documentclass[twocolumn,amsmath,amssymb,aps,preprintnumbers,showpacs,superscriptaddress]{revtex4}

\usepackage{graphicx}
\usepackage{dcolumn}
\usepackage{bm}

\newcommand{\lapprox}{{\scriptscriptstyle\stackrel{<}{\sim}}}

\begin{document}


\title{Commensurability effects in superconducting
Nb films with quasiperiodic pinning arrays}

\author{M.~Kemmler}
\author{C.~G\"{u}rlich}%
\author{A.~Sterck}%
\author{H.~P\"{o}hler}%
\affiliation{%
Physikalisches Institut -- Experimentalphysik II, Universit\"{a}t
T\"{u}bingen, Auf der Morgenstelle 14, D-72076 T\"{u}bingen, Germany
}%
\author{M.~Neuhaus}
\author{M.~Siegel}
\affiliation{IMS, Universit\"{a}t Karlsruhe, Hertzstr.~16,
D-76187 Karlsruhe, Germany
}%
\author{R.~Kleiner}
\author{D.~Koelle}
\email{koelle@uni-tuebingen.de}
\affiliation{%
Physikalisches Institut -- Experimentalphysik II, Universit\"{a}t
T\"{u}bingen, Auf der Morgenstelle 14, D-72076 T\"{u}bingen, Germany
}%

\date{\today}

\begin{abstract} 
We study experimentally the critical depinning current
$I_c$ versus applied magnetic field $B$ in Nb thin films
which contain 2D arrays of circular antidots placed on the
nodes of quasiperiodic (QP) fivefold Penrose lattices.
Close to the transition temperature $T_c$ we observe
matching of the vortex lattice with the QP pinning
array, confirming essential features in the $I_c(B)$
patterns as predicted by Misko {\em et al.}
[Phys.~Rev.~Lett. {\bf 95}(2005)].
We find a significant enhancement in $I_c(B)$ for QP
pinning arrays in comparison to $I_c$ in samples with
randomly distributed antidots or no antidots.
\end{abstract}

\pacs{74.25.Qt, 74.25.Sv, 74.70.Ad, 74.78.Na}
%
%
%
%


\maketitle


The formation of Abrikosov vortices in the mixed state of
type-II superconductors \cite{Abrikosov57} and their
arrangement in various types of ''vortex-phases``, ranging
from the ordered, triangular Abrikosov lattice to
disordered phases \cite{Nelson93,Blatter94,Brandt95} has a
strong impact on the electric properties of
superconductors.
Both, in terms of device applications and with respect to the
fundamental physical properties of so-called ''vortex-matter``, the
interaction of vortices with defects, which act as pinning sites,
plays an important role.
Recent progress in the fabrication of nanostructures provided the
possibility to realize superconducting thin films which contain
artificial defects as pinning sites with well-defined size, geometry
and spatial arrangement.
In particular, artificially produced {\em periodic
arrays} of sub-micron holes (antidots)
\cite{Baert95,Moshchalkov98,Castellanos97,Harada96}
and magnetic dots
\cite{Martin97,Morgan98,VanBael99,Villegas03a} as
pinning sites have been intensively investigated
during the last years, to address the fundamental
question how vortex pinning -- and thus the critical
current density $j_c$ in superconductors -- can be
drastically increased.

In this context, it has been shown that a very stable vortex
configuration, and hence an enhancement of the critical current
$I_c$ occurs when the vortex lattice is commensurate with the
underlying periodic pinning array.
This situation occurs in particular at the so-called
first matching field $B_1=\Phi_0/A$, i.e., when the
applied field $B$ corresponds to one flux quantum
$\Phi_0=h/2e$ per unit-cell area $A$ of the pinning
array.
In general, $I_c(B)$ may show a strongly non-monotonic
behavior, with local maxima at matching fields
$B_m=mB_1$ ($m$: integer or a rational number), which
reflects the periodicity of the array of artificial
pinning sites.

As pointed out by Misko {\em et al.} \cite{Misko05},
an enhancement of $I_c$ occurs only for an applied
field close to matching fields, which makes it
desirable to use artificial pinning arrays with many
built-in periods, in order to provide either very many
peaks in $I_c(B)$ or an extremely broad peak in
$I_c(B)$.
Accordingly, Misko {\em et al.} studied analytically and by
numerical simulation vortex pinning by quasiperiodic chains and by
2D pinning arrays, the latter forming a fivefold Penrose lattice
\cite{Suck02}, and they predicted that a Penrose lattice of pinning
sites can provide an enormous enhancement of $I_c$, even compared to
triangular and random pinning arrays.

We note that the discovery of quasicrystals
\cite{Shechtman84} has, until today, stimulated
intensive investigations of a large variety of
artificially generated quasiperiodic systems, such as
semiconductor heterostructures \cite{Merlin85},
optical superlattices \cite{Zhu97}, photonic
quasicrystals
\cite{Notomi04}, atoms in optical potentials
\cite{Guidoni97}, superconducting wire networks
\cite{Behrooz86} and Josephson junction arrays
\cite{Springer87}. The investigation of the static and
dynamic properties of {\em vortex quasicrystals},
including phase transitions which may be tuned by
temperature and magnetic field, is interesting, both
from a practical point of view (regarding
controllability and enhancement of critical currents
in superconductors) and also with respect to our
understanding of fundamental aspects related to the
physics of quasicrystals.

Here, we present results on the experimental investigation
of matching effects in superconducting Nb thin films
containing various types of antidot configurations.
We studied $I_c(B)$ at variable temperature $T$ close
to the superconducting transition temperature $T_c$,
and we compare Penrose lattices with triangular
lattices, with random arrangements of antidots and
with thin films without antidots.
Our experimental results on Penrose arrays confirm
essential features in the $I_c(B)$ patterns as
predicted in \cite{Misko05}.

\begin{figure}[tb]
\center{\includegraphics[width=0.8\columnwidth,clip]{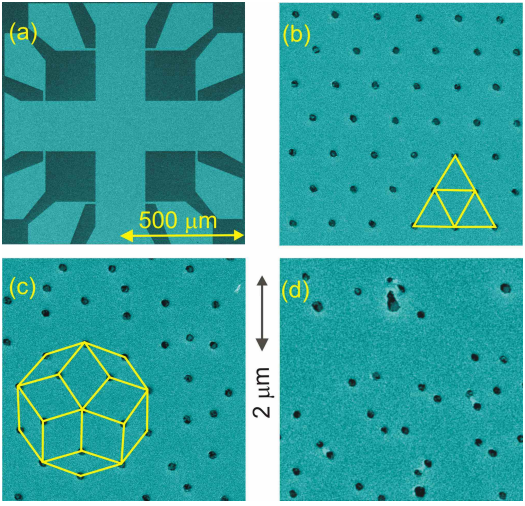}}
\caption{(Color online)
Images of (a) cross-shaped Nb bridge and of different
arrangements of antidots (125\,nm radius): triangular
lattice (b); Penrose lattice (c); random arrangement
(d).
The lines in (b),(c) illustrate the lattice
geometries.
\label{fig:Images}}
\end{figure}

The experiments were carried out on $d=60\,$nm thick
Nb films which were deposited by dc magnetron
sputtering in the same run on five separate
substrates.
Patterning was performed by e-beam lithography and
lift-off to produce cross-shaped Nb bridges with
circular antidots arranged in different geometries.
Fig.~\ref{fig:Images}(a) shows the geometry of the Nb
bridges, with segments of width $w=200\,\mu$m and
length (separation between voltage pads)
$l=600\,\mu$m.
On each chip we are able to directly compare eight different cross
structures.
Six of them contain approximately $N_p=$110,000
circular antidots with radius $r=125\,$nm or 200\,nm,
arranged in either a triangular lattice
[Fig.~\ref{fig:Images}(b)], a Penrose lattice
[Fig.~\ref{fig:Images}(c)], or in a random arrangement
[Fig.~\ref{fig:Images}(d)].
All samples have the same average antidot density
$n_p=0.52\,\mu{\rm m}^{-2}$, which corresponds to a first
matching field $B_1=n_p\Phi_0=1.08\,$mT.
For reference measurements each chip also contains two cross
structures without antidots (''plain`` sample).

The fivefold Penrose lattice consists of two types of
rhombuses with equal sides $a_P=1.54\,\mu$m: ''large``
and ''small`` ones, with angles ($2\Theta$, $3\Theta$)
and ($\Theta$, $4\Theta$), respectively
($\Theta=36^\circ$).
Accordingly, the short diagonal of the small rhombuses
$a_P/\tau=0.952\,\mu$m; $\tau=(1+\sqrt{5})/2\approx
1.618$ is the golden mean.
The rhombuses have been arranged using inflation rules
\cite{Suck02} in order to assure the quasiperiodicity
of the lattice.
%
In the triangular lattice the next-neighbor distance
is $a_T=1.49\,\mu$m.
For the random arrangement of antidots their
$(x,y)$-coordinates were generated by a 2D array
filled with uniform random numbers and then scaled to
give $n_p=0.52\,\mu{\rm m}^{-2}$.

\begin{figure}[tbp]
\center{\includegraphics[width=0.9\columnwidth,clip]{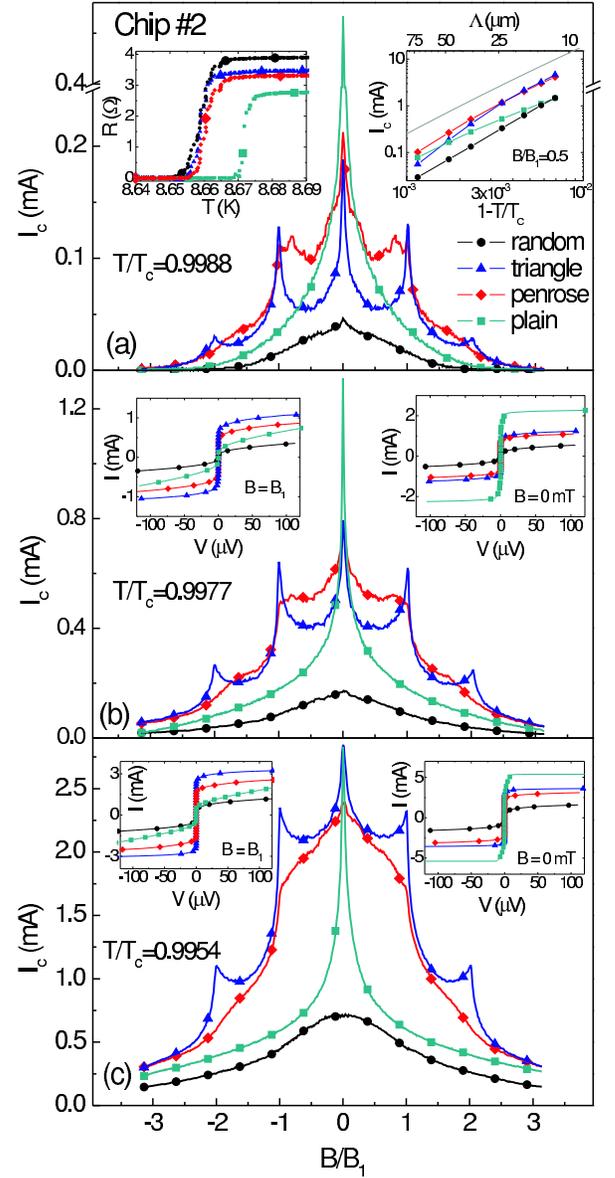}}
\caption{(Color online)
Comparison of four different bridges (chip\#2,
$r=200\,$nm).
Main graphs:
$I_c$ vs $B$ ($V_c=2\,\mu$V); $T$ decreases from (a)
to (c).
Insets in (a):
$R$ vs $T$ at $B=0$, $I=10\mu$A (left) and $I_c$ vs
$(1-T/T_c)$ at $B=B_1/2$ (right); grey solid line is
calculated with a simple core pinning model.
Insets in (b),(c):
$I$ vs $V$ at $B=0$ (right) and $B=B_1$ (left).
\label{fig:IcB-chip2}}
\end{figure}

To characterize our devices, we measured resistance
$R$ vs. $T$ (at $B=0$) and determined $T_c$ of the
different bridges on each chip.
The left inset in Fig.~\ref{fig:IcB-chip2}(a) shows
$R(T)$ curves for perforated bridges and for one
bridge without antidots on chip \#2.
For the perforated bridges, $T_c=8.660\,$K is reduced
by 12\,mK compared to the bridge without antidots.
This can be attributed to a small contamination from
the resist structure during Nb film deposition, as we
observed a similar behavior on other chips.
The perforated bridges also show a larger normal
resistance, which can be ascribed to geometry effects
and to a slight reduction in the mean free path $\ell$
as compared to the plain bridge; again, this is
consistently observed for all chips.
With this respect, the $R(T)$ curves shown in the left
inset of Fig.~\ref{fig:IcB-chip2}(a) are
representative for all devices on various chips which
we investigated.
From the measured resistivity $\rho_{10{\rm
K}}=5.52\,\mu\Omega$cm of the plain film at $T=10\,$K
and the relation $\rho\ell=3.72\cdot
10^{-6}\,\mu\Omega{\rm cm}^2$ \cite{Mayadas72} we
estimate $\ell\approx 6\,$nm.
%
%
%
In total we investigated 12 bridges on three different
chips (\#1, \#2, \#3).
Below we present data for bridges with $r=200\,$nm
antidots on chips \#2 and \#3.
Bridges with $r=125\,$nm antidots on chip \#1 behaved
similarly.

\begin{figure}[tbp]
\center{\includegraphics[width=0.9\columnwidth,clip]{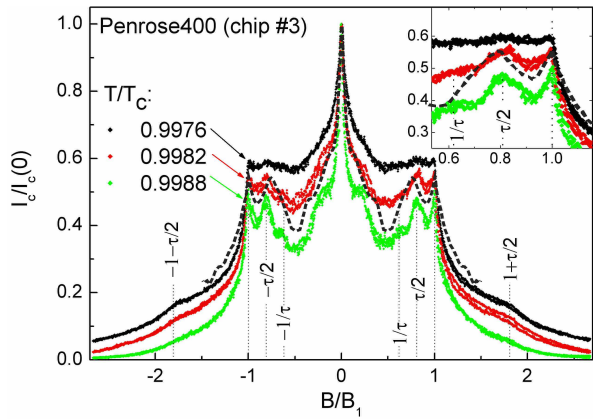}}
\caption{(Color online)
$I_c$ vs $B$ of a Penrose array (chip\#3, $r$=200\,nm,
$V_c=1\,\mu$V) at different $T$ close to
$T_c=8.425\,$K.
$I_c$ is normalized to $I_c(B=0)=0.71\,$mA, 0.39\,mA
and 0.16\,mA at $T/T_c$=0.9976, 0.9982 and 0.9988,
respectively.
Inset shows magnification at $0.5\lapprox B/B_1\lapprox
1.15$.
$B$ has been swept through a full cycle (from its
minimum value up to its maximum and back).
Dotted lines show simulation result redrawn from
Fig.~2(e) in \cite{Misko05} for $N_p=301$.
\label{fig:IcB-Penrose-chip3}}
\end{figure}

Figure \ref{fig:IcB-Penrose-chip3} shows
$I_c(B)$-patterns of a Penrose array at three
different temperatures close to $T_c$.
Here, $I_c$ is defined by a dynamic criterion, i.e.,
via the detection of a threshold voltage $V_c=1\,\mu$V
and the $T$-stability is $\approx 1\,$mK.
For comparison, we insert the calculated $I_c(B)$
dependence (as dashed line), replotted from Fig.~2(e)
in \cite{Misko05} (for $N_p$=301 pinning sites),
without any adjustable parameter.
For these calculations a static $I_c$ definition
(number of pinned vortices over the number of
vortices) was used. However, Misko {\em et al.}
\cite{Misko05} mention that dynamic simulations give
essentially equal results.
We find very nice agreement with our experimental data for
the two highest temperatures (lower and middle traces):
The narrow peak in $I_c(B)$ at $B\approx 0$ broadens
above $B/B_1\approx 0.2$, until a local minimum in
$I_c(B)$ is reached at $B/B_1\approx 0.5$; a two-peak
structure appears with $I_c$ maxima at
$B/B_1=0.81\approx\tau/2$ (broader peak) and at
$B=B_1$ (narrow peak).
Above $B_1$, $I_c$ drops rapidly with increasing $B$.
The two-peak structure was predicted in \cite{Misko05}
for the case that the gain in pinning energy $E_{pin}$
exceeds the increase in elastic energy $E_{el}$
associated with the deformation of a triangular vortex
lattice.
Misko {\em et al.} \cite{Misko05} find the broader
peak at a value of $B_{v/t}/B_1=0.757$, which
corresponds to filling of only three out of four of
the pinning sites on the vertices of the small
rhombuses, except for the $N_d$ cases when two small
rhombuses are connected along one side, and share one
such vacancy.
We find for our large Penrose arrays $N_d=N_s/\tau^2$
($N_s$ and $N_l$ are the number of small and large
rhombuses, respectively).
With the number of vertices in the Penrose lattice
$N_p=\tau N_l=\tau^2 N_s$, one obtains
$B_{v/t}/B_1=2/\tau^2\approx0.764$.
This is just slightly below the matching peak which we
observe at $\tau/2=(N_p-\frac{N_s}{2})/N_p$.
The latter corresponds, e.g., to a vortex
configuration with {\em every second} small rhombus
having only three out of four pinning sites occupied.
In addition, we find at the highest temperature
$T/T_c=0.9988$ a third peak in $I_c(B)$ at
$B/B_1=0.65\approx 1/\tau$, which can be associated
with filling of only three out of four pinning sites
on the vertices of {\em all} small rhombuses.

With decreasing $T$ another peak-like structure
appears at $B/B_1=1.77$, close to $1+(\tau/2)$.
In the same $T$-range, the triangular array with same
antidot size (on the same chip) develops a clear peak
in $I_c(B)$ at the second matching field $B_2$.
This can be explained by an increase of the saturation
number\cite{Mkrtchyan72} $n_s\approx r/2\xi(T)$ above
a value of two ($\xi$ is the coherence length).
I.e., by lowering $T$, pinning of multiquanta occurs.
The observed matching peak at $1+(\tau/2)$ corresponds
then, e.g., to double occupancy of antidots, except
for one antidot on every second small rhombus, which
is only occupied by a single vortex.
The fact that this matching peak is not very
pronounced indicates that $E_{pin}$ only slightly
exceeds $E_{el}$, which also explains the missing of a
matching peak at $B=B_2$.

The matching peaks in Fig.~\ref{fig:IcB-Penrose-chip3}
are most pronounced at the highest temperature,
$T/T_c=0.9988$, and gradually transform into a
plateau-like $I_c(B)$ pattern (within $0.5\le B/B_1\le
1$) at $T/T_c=0.9976$. This plateau is a remarkable
feature, which we associate with the effectiveness of
the many built-in periods of the Penrose lattice.
We note that the Pearl length \cite{Pearl64}
$\Lambda(T)\propto(1-T/T_c)^{-1}$, which sets the
vortex interaction range, changes from $\approx 80$ to
$40\,\mu$m with decreasing $T$ in
Fig.~\ref{fig:IcB-Penrose-chip3}.
In contrast to the calculations in \cite{Misko05},
with a penetration depth of the order of the pinning
lattice spacing $a$, we find distinct matching peaks
only for $\Lambda\gg a$, i.e. if the vortices interact
over a large fraction of the Penrose array.
Certainly, a better understanding of our results
requires a more detailed theoretical analysis,
including the $T$-dependence of the vortex interaction
range, pinning strength and pinning range, as well as
thermal fluctuations, which might be important, in
particular close to $T_c$.


For direct comparison, we simultaneously measured on a
single chip (\#2) three perforated bridges with
different antidot configurations (Penrose, triangular
and random; $r=200\,$nm) and one plain bridge.
Figure \ref{fig:IcB-chip2} shows $I_c(B)$ at three
different values of $T/T_c$.
In contrast to the 'plain' and 'random' samples the
Penrose and triangular arrays show clear matching
effects, with identical $B_1$, as designed.
The triangular array shows very pronounced matching
peaks at $B_1$ and $B_2$.
We do not observe higher order matching peaks, which
indicates that $n_s\approx 2$.

The $I_c(B)$ pattern for the Penrose array is very
similar to the one on chip\#3 with same antidot size
(c.f.~Fig.~\ref{fig:IcB-Penrose-chip3}).
Decreasing $T$, the multiple peak structure at $B\le
B_1$ turns into a broad shoulder.
When $T$ is lowered further [see
Fig.~\ref{fig:IcB-chip2}(c)], the shoulder transforms
into a dome-like structure.
We cannot give a concise explanation for the shape of
this very broad central peak; however, we note that we
observed this on all investigated Penrose arrays.

Comparing absolute values of $I_c$ for the Penrose and
triangular array shows that very close to $T_c$
critical currents at $B=0$, $B_1$ and $B_2$ are quite
similar; however, due to the stronger reduction in
$I_c$ of the triangular array between matching fields,
the Penrose array is superior, in particular for small
fields, below $B_1$ [see Fig.~\ref{fig:IcB-chip2}(a)].
This situation changes with decreasing $T$, as the
reduction in $I_c$ of the triangular array between
matching fields becomes much less pronounced [see
Fig.~\ref{fig:IcB-chip2}(c)].
I.~e., we cannot confirm the prediction \cite{Misko05}
that the Penrose lattice provides an enormous
enhancement in $I_c(B)$ over the triangular one,
except for $T$ very close to $T_c$ and fields between
$B=0$ and the first matching field.
This is also visible in the right inset in
Fig.~\ref{fig:IcB-chip2}(a), which shows
$I_c(1-\frac{T}{T_c})$ at $B=0.5B_1$ for all four
samples.
Above $1-\frac{T}{T_c}\approx 3\cdot 10^{-3}$ the
Penrose and triangular array show almost the same
$I_c(T)$, which approaches, within a factor of two,
the Ginzburg-Landau depairing current $I_{dp}\propto
H_c^2\xi\propto(1-\frac{T}{T_c})^{3/2}$, with the
thermodynamic critical field $H_c$.
This scaling can be derived within a simple core
pinning model, neglecting vortex-vortex interactions.
The resulting $I_c(T)$ -- shown as grey line in the
right inset in Fig.~\ref{fig:IcB-chip2}(a) -- is then
simply determined by the maximum pinning force of an
antidot, which we calculated, following the approach
in \cite{Blatter94} based on the London approximation
\cite{Mkrtchyan72}.
The 'random' sample gives always significantly smaller
$I_c$ as the triangular and Penrose arrays.
However, it shows a steeper $I_c(T)$-dependence, which
scales as $(1-\frac{T}{T_c})^q$ over the investigated
range of $T$, with an increase in $q$ from $q=2$ at
$B=0$ to $q=2.2$ at $B=0.5B_1$ to $q=2.4$ at $B=B_1$.
This behavior cannot be explained within our simple
model and certainly deserves further investigations.

Finally, we note that the choice of the voltage
criterion $V_c$ for determining $I_c$ does not
significantly affect the shape of the $I_c(B)$-curves;
we checked this for various samples and $V_c$ ranging
from 10\,nV to 10\,$\mu$V.
For a more detailed analysis, we measured current
voltage characteristics (IVCs) as shown in the insets
of Fig.~\ref{fig:IcB-chip2}(b),(c) for all four
bridges at $T/T_c$=0.9977 and 0.9954, respectively.
The IVCs of the perforated bridges evolve smoothly
into the resistive state.
In particular, there is no sudden change in
differential resistance $R_d$ pointing to a sudden
increase of the number of moving vortices.
At $B=0$ (right hand insets) the IVCs of the different
bridges do not cross (this holds up to at least 1 mV).
Such a crossing is, however, observed in finite
fields.
As shown in the left hand insets ($B=B_1$), $R_d$ of
the plain bridge is smaller than for the other
bridges.
At $V\approx300\,\mu\rm V$ its IVC intersects those of
the triangular and Penrose arrays.
A similar crossing is also observed for other values of
field and temperatures, however, always well above the
threshold voltage $V_c$ used for determining $I_c$.


In conclusion, we experimentally verified theoretical
predictions of the main features of the critical
current dependence on the applied magnetic field in a
superconducting thin film with a quasiperiodic Penrose
lattice of antidots at temperatures close to $T_c$.
In particular, we found matching of the vortex lattice with
the quasiperiodic lattice of pinning sites very close to
$T_c$, and we associate various matching peaks in $I_c(B)$
with distinct regular arrangements of the vortices.
In addition, we directly compared different arrangements of
artificial pinning sites in our Nb films.
We find an enhancement of $I_c$ in films with Penrose
lattices as compared to films with random arrangement
of pinning sites and films without artificial pinning
sites, however no significant enhancement of $I_c$ as
compared to films with triangular antidot lattices.
With respect to applications, it will be interesting
to perform more detailed investigations on the effect
of optimum antidot size (i.e.~saturation number) and
density over a wide range of temperatures in
quasiperiodic pinning arrays.

We thank Eric Sassier for his support on the
measurement setup, and we gratefully acknowledge
helpful discussions with V.~Misko, F.~Nori and
A.~V.~Silhanek.
This work was supported by the Deutsche
Forschungsgemeinschaft (DFG; KL-930/10 and SFB/TR21).
M.~Kemmler gratefully acknowledges support from the Evangelisches
Studienwerk e.V.~Villigst.

{\em Note added.} After first submission of our
manuscript we learned about related work by Villegas
{\em et al.}\cite{Villegas06} and Silhanek {\em et
al.}\cite{Silhanek06} on quasiperiodic pinning arrays
in superconducting films with magnetic dots.

\bibliography{vmatt,penrose,QP-Crystals}

\end{document}